\documentclass{jdmdh}
\usepackage[utf8]{inputenc}
\usepackage[english]{babel}
\usepackage{comment}
\usepackage{array}
\usepackage{pgfplots}
\usepackage{graphicx}
\usepackage{subfigure}
\usepackage{listings}
\usepackage{color}
\usepackage{dirtree}
\usepackage{csquotes}
\usepackage{tikz}

\pgfplotsset{compat=1.17}
\titlespacing*{\section}
{0pt}{2.5ex plus 1ex minus .2ex}{1.3ex plus .2ex}
\titlespacing*{\subsection}
{0pt}{1.0ex plus 1ex minus .2ex}{1.3ex plus .2ex}
\titlespacing*{\subsubsection}
{0pt}{1.0ex plus 1ex minus .2ex}{1.3ex plus .2ex}

\usepackage{natbib}
\usepackage{blindtext}
\usepackage{hyperref}

\title{From Books to Knowledge Graphs}

\author[1]{Natallia Kokash\thanks{Corresponding author: \href{mailto:n.kokash@uva.nl}{n.kokash@uva.nl}.}} 
\author[2]{Matteo Romanello}
\author[3]{Ernest Suyver}
\author[1]{Giovanni Colavizza}
\affil[1]{University of Amsterdam, Amsterdam, The Netherlands} 
\affil[2]{University of Lausanne, Lausanne, Switzerland} 
\affil[3]{Brill, Leiden, The Netherlands} 

\begin{document}

\maketitle

\abstract

The digital transformation of the scientific publishing industry has led to dramatic improvements in content discoverability and information analytics. Unfortunately, these improvements have not been uniform across research areas. The scientific literature in the arts, humanities and social sciences (AHSS) still lags behind, in part due to the scale of analog backlogs, the persisting importance of national languages, and a publisher ecosystem made of many, small or medium enterprises. We propose a bottom-up approach to support publishers in creating and maintaining their own publication knowledge graphs in the open domain. We do so by releasing a pipeline able to extract structured information from the bibliographies and indexes of AHSS publications, disambiguate, normalize and export it as linked data. We test the proposed pipeline on Brill's Classics collection, and release an implementation in open source for further use and improvement. 



\section{Introduction}
\label{sect:Introduction}

The scientific publishing industry has transitioned to information analytics. Researchers immensely benefit from the indexed content and provision of advanced search engines, primarily based on scientific publication metadata, citations and, increasingly, contents. Notable examples include Google Scholar, Semantic Scholar, Dimensions, and more~\citep{martin-martin_google_2021,visser_large-scale_2021}. Advances in the adoption of the open science agenda have resulted in the increased availability of open and structured scientific literature data, for example via PubMed~\citep{white_pubmed_2020}, OpenAlex\footnote{\url{https://openalex.org}.} and OpenCitations~\citep{peroni_opencitations_2020}. Unfortunately, this digital transformation is not occurring uniformly in all research areas. A divide persists for the Arts, Humanities and, to a lesser degree, the Social Sciences (AHSS): the indexation of this literature is still lagging, in particular with respect to historical backlogs and publications in languages other than English~\citep{colavizza_citation_2019}, both essential parts of it~\citep{kulczycki_publication_2018,kulczycki_multilingual_2020}. Recently, proposals have been made to better support LAM organisations (Libraries, Archives, Museums) in the indexation of the literature they collect~\citep{colavizza_references_2018,colavizza_case_2021}. Another obstacle to a better indexation of AHSS literature is that the publishing ecosystem supporting these disciplines consists of many specialized organizations that cannot, individually, transform established practices and develop the large-scale services required for systematic indexation. We address this latter challenge here.

We propose to start bridging the gap by supporting small and medium AHSS publishers to create and maintain their own \textit{knowledge graphs} (KG): the technology underpinning modern scientific search engines~\citep{luan_multi-task_2018,jaradeh_open_2019}. We do so by releasing an end-to-end information extraction pipeline which takes unstructured publications as input and outputs structured, relational information. The proposed pipeline extracts the relevant information, performs entity linking to equip them with identifiers, and structures information following the SPAR and OpenCitations data models~\citep{vrandecic_spar_2018,pan_opencitations_2020}. The resulting linked data is made ready for ingestion into OpenCitations. AHSS literature, composed to a significant degree of books, already contains rich and well-structured information -- such as back matter contents including references and indexes -- which we leverage, automatically mine and interlink~\citep{reitz_experiments_2019}. In openly releasing this pipeline, our aim is to foster its adoption and future extension by AHSS publishers, and contribute to a better indexation of AHSS literature in scientific search engines. 

Our work has been conducted in collaboration with Brill, a leading Humanities publisher. The Brill Classics catalogue has been used as a case study for the development of the proposed pipeline. The rest of this paper is structured as follows: Section \ref{sect:content} presents the challenge of extracting structured information from AHSS publications, Section \ref{sect:pipeline} presents our proposed pipeline, Section \ref{sect:graph} discusses the resulting Knowledge Graph, while Sections \ref{sect:evaluation} and \ref{sect:related-work} conclude with an evaluation of the pipeline and the discussion of related literature.

\section{Content representation}
\label{sect:content}

In this section we describe the structure and format of AHSS publications, and the Brill's corpus we used as a case study.

\subsection{The structure of AHSS publications}
\label{sect:content-ahss}
\vspace{1em}

An AHSS book (monograph or edited volume) typically consists of \emph{front matter}, \emph{body}, and \emph{back matter}. 
The \emph{front matter} usually contains a title page, copyright page, table of contents, and one or more  personal preface or introductions. 
The \emph{body} or main text typically consist of a number of chapters or book sections, among which the first (introduction or prologue) and the last (conclusions or epilogue) stand out as they open the narrative and sum up the core ideas. 
The \emph{back matter} may contain appendices, indexes and a bibliography. These supplementary sections are meant to inform the reader about certain aspects of the content, and the choices largely depend on each particular book's design.

The front and back matter are the parts of a book we are most interested in the context of building a KG that links a given publication with related works and subjects. The front matter provides metadata necessary to identify a publication: author, title, publication year, and (for modern works ) persistent identifiers like ISBN (International Standard Book Number) or DOI (Digital Object Identifier).

The analysis of the back matter of published literature in the AHSS domain was the most challenging aspect we dealt with due to the variety of its content, formats, conventions, and organization rules. Two types of the back matter sections particularly relevant to us are:
\begin{itemize}
    \item \emph{Bibliography}: a list of references, usually to secondary literature. Primary sources are, instead, often listed in an index.
    \item \emph{Index}: a curated list of relevant items and their mentions in the book.
\end{itemize}

\begin{figure}
    \centering
    \includegraphics[width=0.85\textwidth]{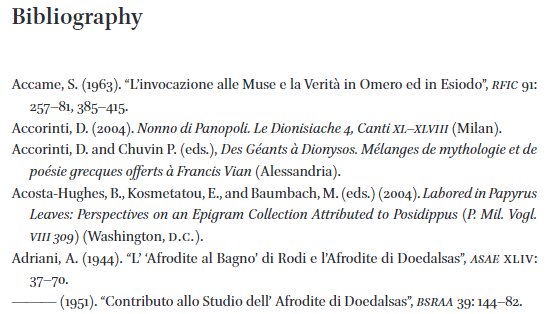}
    \caption{Bibliography references (DOI: \href{https://doi.org/10.1163/9789004289598_008}{10.1163/9789004289598\_008})} 
    \label{fig:bm-bib} 
\end{figure}

\begin{figure}
    \centering
    \subfigure[\scriptsize DOI: \href{https://doi.org/10.1163/9789004289598_010}{10.1163/9789004289598\_010}]{
         \includegraphics[width=0.4\textwidth]{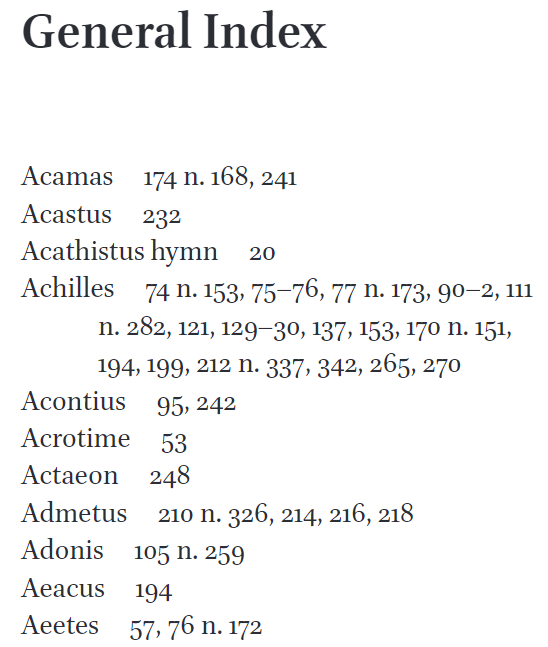}
         \label{fig:bm-idx-general}
    }
    \subfigure[\scriptsize DOI: \href{https://doi.org/10.1163/9789004289598_009}{10.1163/9789004289598\_009}]{
         \includegraphics[width=0.49\textwidth]{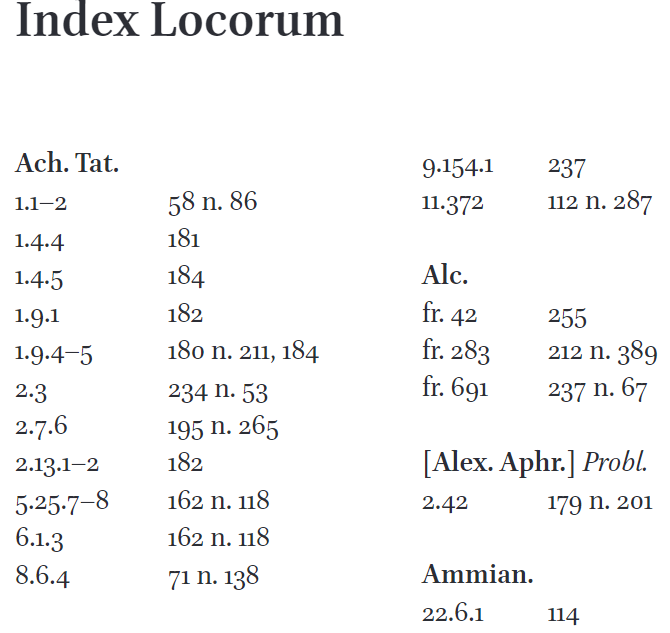}
         \label{fig:bm-idx-locorum}
    }
    \caption{Index references} 
    \label{fig:bm-index}
\end{figure}

If the book is a \emph{monograph}, i.e., written by one author, the bibliography tends to be at the back of the book. If the book is an \emph{edited volume}, i.e., it contains chapters written by different authors, the bibliography tends to be at the end of each chapter. 
Indexes are usually found at the back of the book, but not all books have them. Index types may vary: sometimes there is only one, sometimes, there are several. A usual pattern is an index of names (\emph{index nominum}), an index of places (\emph{index locorum})\footnote{Note that an index of places is not about place names (geolocations), but about passages in ancients texts (the so-called \emph{loci}).} and an index of things, i.e., subjects (\emph{index rerum}). Depending on the subject, there can be more specialized indexes.

Figure~\ref{fig:bm-bib} shows a fragment of back matter with bibliography references. Figure~\ref{fig:bm-index} shows examples of index sections: an index of general terms (Figure~\ref{fig:bm-idx-general}) and an index of places (Figure~\ref{fig:bm-idx-locorum}). 

\subsubsection{Citations and bibliographic references}
References in bibliographies must be distinguished from citations in the text. A citation is an often-abbreviated pointer to a location in a publication. The book is usually identified by a combination of author name and publication year. The reference does not give the location, but instead gives the complete bibliographical data for that publication. For example: on p. 13, footnote 63 of our example book~\cite{StudiesinColluthusiAbductionofHeleni}, we find this citation: 
\begin{displayquote}
\small 
    Miguélez-Cavero 2008, 87.
\end{displayquote}

This corresponds to the following reference in the bibliography on page 292:
\begin{displayquote}
\small
    ——— (2008). Poems in Context: Greek Poetry in the Egyptian Thebaid 200–600 AD (Berlin).
\end{displayquote}
Note that the author name is replaced by three dashes. This is because the author name is already mentioned in a previous reference. This is a matter of citation style: many such styles exist. Most Brill book series follow the Chicago Author-Date style.~\footnote{\url{https://www.chicagomanualofstyle.org}.}

\subsubsection{Indexes}

The first two items in the \emph{general index} of the same example read:
\begin{displayquote}
\small 
    Acamas 174 n. 168, 241 \\
    Acastus 232
\end{displayquote}

This is a two-column table: on the left are the index terms, on the right are the page numbers (locations in the book).
Multiple locations are separated by a comma, the left and right parts are separated by a tab. The presence of `n.' in the location part indicates that the index entry refers to a footnote. To save space, such two-column definitions are often arranged into several columns on every page. 

The \emph{index locorum} tells us which passage (\emph{locus}), from which ancient work occurs on which page of the work in question. For example, the following index entry
\begin{displayquote}
\small 
  Alc. $\quad$ fr. 42 $\,$ 255  
  $\quad$ fr. 283 $\,$ 212 n. 389 
  $\quad$ fr. 691 $\,$ 237 n. 67
\end{displayquote}

refers to the name of the author, the poet \emph{Alcaeus}, in abbreviated
form, on the left, and a set of his fragment occurrences, on the right. No specific work of his is mentioned in the book, only fragments (hence, abbreviation \emph{fr.}). In the cases where we do have the works, they are commonly mentioned in an abbreviated form too: 
\begin{displayquote}
\small 
    Aesch. Ag. 681–98 $\,$ 82
\end{displayquote}


\subsection{Structured annotations} 
\label{sect:content-jats}

In order to extract useful data from the publication sources, we first need to locate its relevant parts. If the work were given in a single file (e.g., PDF), the only option would be to analyze the text and the layout of the file to distinguish front matter, body, and back matter. However, the essential metadata such as author's name, title, tables of contents, etc., is part of the publication and dissemination processes and is normally kept by publishers in a structured form.

The \emph{Journal Article Tag Suite}~\cite{JATS:2021} is a standard XML vocabulary designed to model current journal articles. It provides a named collection of XML elements and attributes that can be used to mark the structure and semantics of a single journal article. Originally, JATS was used for STEM (Scientific, Technical, Engineering and Medical) articles, but now publishers in AHSS also widely adopt this markup language. Tools have been developed to export JATS out of many other formats, including Microsoft Word and LaTeX, and to generate  quality PDF, HTML, and various eBook formats out of it.

\lstset{
    basicstyle=\ttfamily\footnotesize
}

Book collections are often annotated using a JATS extension called \emph{Book Interchange Tag Set}~\cite{BITS:2016}. The intent of BITS is to provide a common format for publishers and archives to exchange book content. The tag set describes both the metadata and the narrative content of a book and its components, as well as collection-level metadata for book sets and series. The BITS annotation may include the following (optional) components:
\begin{itemize}
    \item \emph{Collection metadata}: the \lstinline|<collection-meta>| element describes book sets or series to which this book or book part belongs. 
    \item \emph{Book metadata}: the \lstinline|<book-meta>| element includes nested tags defining, for example, the title (\lstinline|<book-title-group>|), the contributors (\lstinline|<contrib-group>|), the date of publication (\lstinline|<pub-date>|), the publisher (\lstinline|<publisher>|), etc. 
    \item \emph{Front matter}: the \lstinline|<front-matter>| element provides the textual front material for a book, such as a dedication, foreword, or preface.
    \item \emph{Body of the book}: the \lstinline|<book-body>| element defines the structure of its main textual and graphic content. The body of a book contains book parts \lstinline|<book-part>| (which may be called parts, sections, chapters, modules, lessons, or whatever divisions a publisher uses). Book parts are recursive and may contain other book parts.
    \item \emph{Back matter}: the \lstinline|<book-back>| element contains the ancillary information such as lists of references (\lstinline|<ref-list>|) and indexes (\lstinline|<index-group>|).
\end{itemize}

Although the JATS/BITS format provides tags for bibliography and index representation, in our case study, these data is not available in this structured form and must be retrieved from PDF files. 

\subsection{Brill's case study}
\label{sect:content-brill}

The dataset used for the creation of our pipeline and a KG comprised of books in the field of Classics. This field was selected because of the Brill's domain expertise and experience with information extraction of this type of content. We chose books over other publication types because they are characteristic of the AHSS and because, unlike journal articles, they have extensive back-of-the-book indexes that can be used for content discovery.

The corpus consists of 1816 books with an average of 369 pages. The books were produced in the period 2006-2021. 
The books included 965 edited volumes (collections of chapters by different authors), and 851 monographs (books on one subject written by a single author).

At Brill, books are archived in the following manner. A folder is created using the ISBN for the online version (`e-ISBN'). It contains content and metadata and is archived in a compressed form. The content is usually in the form of a single PDF for the entire book, as well as PDFs for the individual book sections, and metadata are in BITS XML, for example: 
\dirtree{%
.1 9789004339460\_BITS.zip.
.2 9789004339460\_webready\_content\_s001.pdf.
.2 ....
.2 9789004339460\_webready\_content\_s018.pdf.
.2 9789004339460\_webready\_content\_text.pdf.
.2 9789004339460\_webready\_content\_text.xml.
}

A large obstacle, characteristic for small-to-medium-sized publishers, is the lack of consistency in the typography and the content due to the following factors: 

\begin{itemize}
    \item \emph{Content variability}. Books have different subjects; authors belong to different fields;  countries, even institutions, have their own conventions. Brill books usually belong to series, and each series has its own typographical, orthographic, and other conventions. Within the series, volumes may differ. Moreover, the entire workflow (writing, peer review, copy editing, indexing) is manual and lacks the use of standardized tools.
    \item \emph{Choice of typesetters}. Books are produced by different typesetters, e.g., using InDesign, PageMaker, or LaTeX. Until 2018, there was no Brill Typographical Style (BTS)~\citep{Brill:2020}. Even now, BTS, despite its name, is not a uniform typographical style applied to all Brill content, but a limited set of instructions. The execution of BTS is left to the typesetters. BTS allows the idiosyncrasies of the individual series to continue, e.g., BTS does not prescribe a citation style.
    \item \emph{Publishing process evolution}. In 2006, the number of books produced was less then a third of what is now~\citep{vanDerVeen2008}. A large number of small typesetters were used, and author PDFs were common. As time went by and the volume of content increased, more uniformity was imposed on the workflow. In particular, tools were introduced for submission, ERP (Enterprise Resource Planning), workflow, and typesetting, including the Brill typeface~\cite{brillTypeface}. 
 \end{itemize}

Manuscripts are mostly submitted in the form of MS Word documents. Authors are responsible for the creation of the citations and references. In most cases, they are also responsible for the creation of indexes, and rarely use any specialized tools for this. Ultimately, all typographical information is stripped by the typesetter and rebuilt using their own systems. The typeset manuscript is returned to the author for proofreading. A number of proofs (usually two or three) is submitted and returned in what is still a manual process. The final result is a carefully proofread and typeset PDF, published in print and online. 
The same PDF is archived and distributed to discovery services, together with its metadata.

The catalogue we used reflects the challenges Brill faces on the way of transitioning to information analytics, and that these challenges are characteristic of other AHSS publications and publishers.

\section{Knowledge retrieval pipeline}
\label{sect:pipeline}

\begin{figure}
    \centering
    \includegraphics[width=0.9\textwidth]{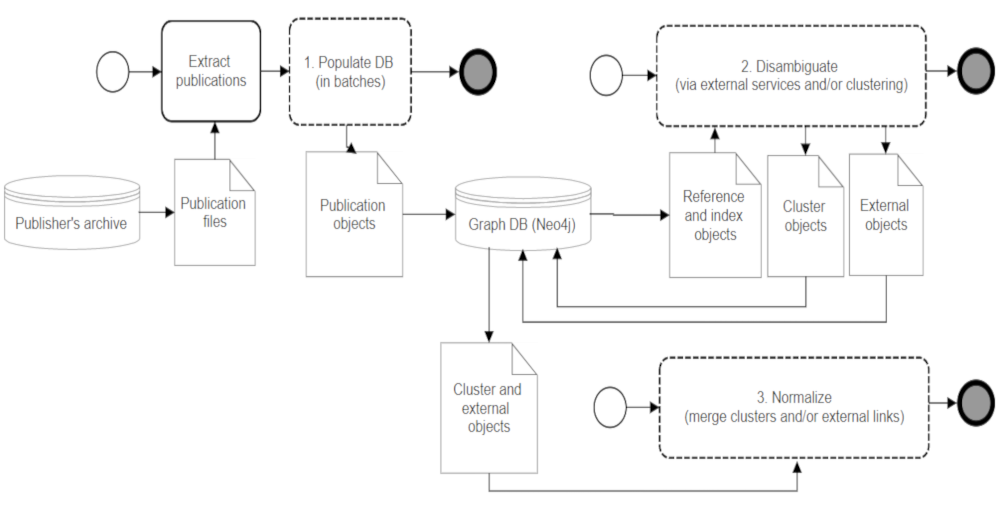}
    \caption{Knowledge graph construction process} 
    \label{fig:pipeline-overview}
\end{figure}

Keeping in mind the organizational structures and assumptions outlined in the previous section, we start our knowledge graph construction from locating and analyzing relevant publication parts. The main steps of our data processing pipeline are illustrated in Figure~\ref{fig:pipeline-overview}. The figure shows three sub-processes, each demarcated by start (white circle) and end (grey circle) events, and data exchange between them.   

The first process is dedicated to data extraction and database population. At this stage, we access a publisher's catalogue and analyze the structural content of individual publications in it. Since such catalogues can be very large, we assume the need to process data in batches of manageable sizes. The batches can be processed sequentially or in parallel, and the process can be stopped and resumed later in time. It is also useful to be able to redo the extraction of data from selected publications without the need to run the whole pipeline again. Given a publication archive or folder, we extract essential data from its metadata file, front and back matter, and create a structured publication object that then gets serialized and stored in a graph database. The graph-based representation of the publication consists of a node that describes its metadata and a set of nodes representing related entities: contributors, identifiers, and, most significantly, bibliographic references and index objects. 

The second stage is dedicated to the disambiguation of bibliographic references and index objects created at the step above. This process covers two major procedures: 
\begin{itemize}
    \item firstly, we identify and cluster bibliographic references that correspond to the same published work version and index terms that refer to the same thing (e.g., subject, person).    
    \item Secondly, we match the extracted reference objects with objects representing the same entities in available public authority records. 
\end{itemize}
The output of this stage consists of a set of new graph nodes representing reference clusters or external links related to the bibliographic references or index objects in our database.

Finally, at the third stage, a number of normalization procedures can be performed to merge cluster and external link nodes corresponding to the same reference or index term. Such node duplicates originate from the fact that each batch is processed independently, i.e., we always create necessary graph nodes while in practice it may be possible to link newly added bibliographic reference and index nodes to existing cluster and external link nodes. 

\subsection{Extracting data from PDF files}
\label{sect:pipeline-pdf}

For each publication folder, we start from locating the JATS/BITS XML file and extracting metadata identifying the publication: title, contributors (authors and/or editors), publication year, publisher, DOI and ISBN identifiers. Then we search for book parts that contain bibliographic references and index lists. These parts are not marked in any special way in our sample archives, so we rely on keywords \emph{bibliography} and \emph{index} in the book part title to locate the corresponding PDF files.

Additionally, we classify index lists to determine what information they provide. Currently, we recognize the following index types: \emph{verborum} (general), \emph{locorum} (citations), \emph{nominum} (names, ancient and modern), \emph{rerum} (subjects), \emph{geographicus} (geographic locations), \emph{bibliographicus} (manuscripts), \emph{museum} (museums), and \emph{epigraphic} (inscriptions). The classifier relies on a number of hits between the index title and predefined terms commonly used to define such indexes. For example, an index file is likely to be an \emph{index nominum} (index of names) if any of the following words is mentioned in its title: `nominum', `nominvm', `propriorvm', `name', `proper', `person', `personal', `people', `writer', `poet', `author', etc. Also, similar terms in other languages are considered: `auteur', `eigennamen', `noms', `propres', `personnages', etc. Furthermore, terms like `ancient', `antique', `classical', `medieval', `greek', `egyptian', `latin', etc. indicate that this is an index of ancient names. The keyword lists are manually created by analyzing titles from a subset of our case study corpus. Some publications may contain highly specialized indices which we cannot properly classify and handle as general. The index types may help parsing index reference text into meaningful parts (i.e., understand that the first word in an entry like \emph{Cicero Att. 1.16.1} is an author name, the next term is their work, and the following numbers are loci) and to help with term disambiguation (e.g., one may want to use a specialized service such as the Classical Works Knowledge Base~\cite{cwkb:2020} for linking this entry to a certain passage by M. Tullius Cicero from `Epistulae ad Atticum'~\footnote{\url{http://www.perseus.tufts.edu/hopper/text?doc=urn:cts:latinLit:phi0474.phi057.perseus-lat1:1.16.1}}). 
\begin{figure}
    \centering
    \includegraphics[width=1\textwidth]{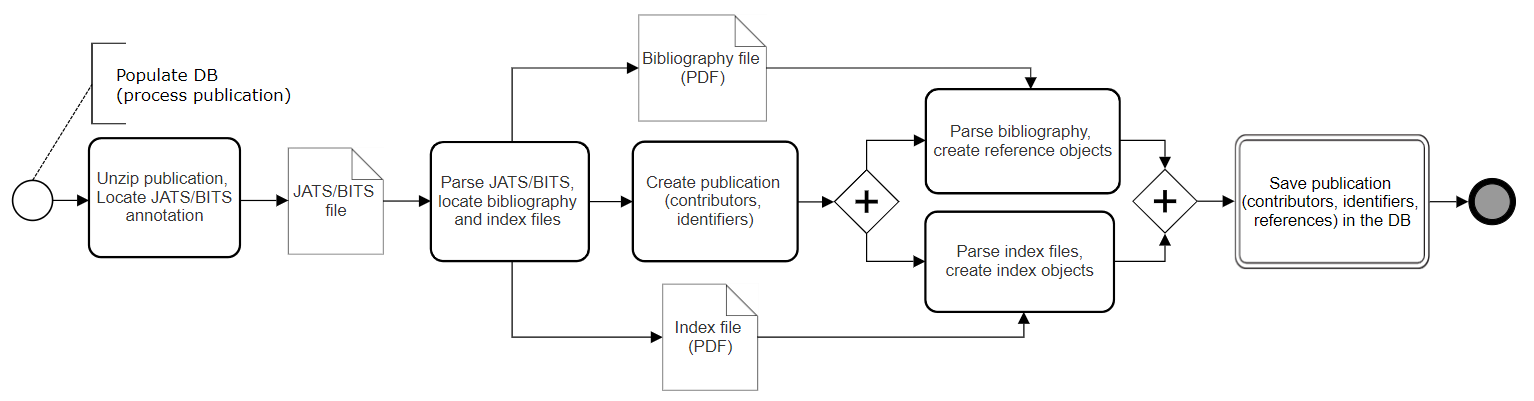}
    \caption{Data extraction pipeline.} 
    \label{fig:pipeline-populate}
\end{figure}

Figure~\ref{fig:pipeline-populate} sketches the process of data extraction for the initial KG construction. After locating the bibliography and index files, the key tasks are to split these PDF files into lists of individual bibliographic and index references. Each item from the lists of textual references  
will then produce structured reference objects serialized as KG nodes.

As was pointed out in Section~\ref{sect:content}, the syntax and the layout of bibliography and index files are rather versatile, there are no standardised ways to represent these data. Even when a popular format such as CMS Author-Date is used across a publication, splitting the bulk data into individual references is still a daunting task: references can be placed in one or several columns, their parts can be separated by commas, different units of indent, alignment, long references can spill into several lines or have nested structure with several sub-levels and relation to above content. However, it can be observed that both bibliographic and index references are \emph{lists}, and, hence, are usually aligned in a uniform way, i.e., each new entry starts from the same horizontal offset as others. This observation also holds for multi-line and multi-level definitions: references that do not fit into one line, resume on the next line, typically, with some extra offset. The same observation is valid for the second column in publications with double-column layout and multi-column index files. 

Our main tool for data retrieval is PDFMiner~\cite{pdfMiner}, a text extraction tool for PDF documents which also obtains the exact location of the text as well as other layout information (e.g., font size). PDFs drastically differ from usual text representation formats -- it is primarily a graphic representation and contains, essentially, instructions on how to place objects on a display or paper. In most cases, it has no logical structures such as sentences or paragraphs. However, PDFMiner provides the class {\texttt{PDFPageInterpreter}} that gives access to the page content. Further, we iterate over the {\texttt{LTTextContainer}} elements of the extracted pages, and, given coordinates of the text containing boxes, build a map that relates each unique $x$-coordinate position with the number of lines starting from it. This is done separately for odd and even pages. Given such a map and assuming that references constitute the majority of the information on pages from a dedicated back matter file, we are able to determine one or more horizontal offset positions where new references start, in the following way:
\begin{enumerate}
    \item sort the map in descending order according to the number of lines per offset.
    \item Mark the left-most position with a large number of lines as the starting position for new references. A map entry with a small offset to the right from this position will indicate the continuation of a long reference or a second level of the index reference (similarly, we can establish the presence of the third level for index definitions).
    \item A map entry with a large number of lines and a significant offset (close to the half of the text width) will signify a two-column format. Similarly to the first column, any lines with a small offset from this position represent long reference continuation or index sub-levels.  
    \item Ignore entries with just a few lines per page: they correspond to `noise' such as e.g., title, subtitle, page number, footnote.
\end{enumerate}

\subsubsection{Parsing bibliographic references}

The previous step produces an ordered list of text references per publication, some of which follow the CMS Author-Date format:
\begin{itemize}
\small
   \item Andrewes, A. 1961, ``Philochoros on phratries'', Journal of Hellenic Studies, vol. 81, pp. 1–15.
   \item ——. 1994. ``Legal space in classical Athens'', \textit{Greece and Rome}, vol. 41, pp. 172–186.
   \item ...        
\end{itemize}
While others do not:
\begin{itemize}
\small
   \item Vernant, Jean-Pierre, \textit{Mythe et société en Grèce ancienne} (Paris, 2004).
   \item Vernant, Jean-Pierre, ``One... Two... Three: Eros,” in \textit{Before Sexuality: 
        The Construction of Erotic Experience in the Ancient Greek World}, 
        ed. Donald M. Halperin, John J. Winkler, and Froma I. Zeitlin (Princeton, 1999), 465-478.
    \item ...        
\end{itemize}

However, depending on the purpose of the KG and further processing steps, it may be necessary or desirable to parse such textual definitions into structured objects.
In particular, since publications are commonly identified by the author-date pair, it is useful to extract the author name and publication year, and consider the rest of the reference text to be a title. Listing~\ref{lst:bib-parser} provides a sample implementation of a bibliography reference parser using the Python's Pyparsing library~\cite{pyparsing}: it attempts to retrieve a list of author names (assuming each is  defined as a family name followed by one or several, possibly abbreviated, given names), followed by a year (or a year range), and assume the rest of the line constitutes a title. In this grammar, {\texttt{ppu}} is an abbreviation for the {\texttt{pyparsing\_unicode}} class that defines various useful sets of Unicode characters. Although not every reference text will be accepted by this grammar, the library makes it easy and straightforward to design a suitable parser should there be more details known about the bibliography editing style.    

\begin{lstlisting}[caption={Bibliography reference parsing}, label={lst:bib-parser}, language={Python}]
  dot = Literal(".")
  comma = Literal(",")
  alphas = ppu.Latin1.alphas+ppu.LatinA.alphas+ppu.LatinB.alphas
  family_name = Word(alphas+"-", min=2)
  init_name = Char(alphas)+dot+Optional("-"+Char(alphas)+dot)
  same = Word("-")+dot.suppress()
  year_or_range = r"\d{4}[a-z]?([,-]\d{4})?"
  year = Regex(year_or_range)+Optional(dot|comma).suppress()
  author = family_name("LastName")+comma+OneOrMore(init_name("FirstName"))
  author_list = Combine((author|same)+Optional(dot|comma).suppress()
  bib_reference = author_list("author")+year("year")+restOfLine("title")
\end{lstlisting}

In our generic implementation, we rely on a similar-style pattern to parse bibliography entries in the CMS Author-Date format, with some technical details to account for abbreviations like \emph{eds.} (editors), and on a simple heuristic to parse entries not accepted by such a strict grammar. The latter attempts to find a year or a year range (using a regular expression from line 7 in Listing~\ref{lst:bib-parser}) anywhere in the text. It also assumes that the list of authors is always in front of the reference text, and that the work's title is the longest sentence between a given set of separators, such as dots or quotation marks.         

\subsubsection{Parsing indexes}

As our examples in Section~\ref{sect:content} illustrated, indexes in AHSS publications provide very specialized information under the assumptions that the reader knows how to interpret their content. The extraction of this information requires formal grammars for various index types and various conventions a publisher and/or authors choose to use. In our case study corpus, we observed the repeated use of certain patterns for certain index types, and developed a set of grammars for them. For example, the majority of index locorum entries are composed of a \emph{label} consisting of one or more words, followed by a \emph{loci} which are numeric expressions (possibly, with several levels separated by dots or dashes), and a set of occurrences in the current work, i.e., page numbers. Listing~\ref{lst:idx-parser} provides a sample grammar for parsing strings that would accept index entries like:
\begin{displayquote}
 \small
  Agamemnon 42–4 591, 593 \\
  Aeschylus Agamemnon 203–4/216–17 433 \\
  Aeschylus Agamemnon 6–7 19 14 586 22 232, 410, 619 32 ff. 129 \\
  Aeschines 2.157 291
\end{displayquote}

\UseRawInputEncoding

\begin{lstlisting}[caption={Index parsing}, label={lst:idx-parser}, language={Python}]
  alphas = ppu.Latin1.alphas+ppu.LatinA.alphas 
    +ppu.LatinB.alphas+ppu.Greek.alphas+"\"'.-_:&()/?"
  ocr = delimitedList(Word(ppu.Latin1.nums))
  locus_fragment = Word(ppu.Latin1.nums+".–=")+Optional(oneOf("ff."))
  locus = locus_fragment+Optional("/"+locus_fragment)
  label = OneOrMore(Word(alphas+",;"))
  index = label("label")+OneOrMore(locus("locus")+ocr("occurrences"))
\end{lstlisting}

A variation of the above pattern without \emph{loci} part would accept index entries typical for indexes of names and subjects. 
In practice, one has to design such grammars given a complete list of syntactic rules and special keywords.
On the plus side, the existing grammar specification libraries like the one used in our tool make it easy to design on-demand parsing algorithms and pass them into the processing pipeline whenever it is applied to a certain dataset.    

\subsection{Disambiguation}

\emph{Disambiguation} refers to the removal of ambiguity to narrow down the meaning of words or distinguishing between similar things (names, locations, subjects, etc.). In the context of our application, disambiguation is also an instrument to determine whether differently phrased bibliographic references imply the same published work, and whether index entries refer to the same term or concept despite some syntactic mismatch in their labels.  

To be able to recognize that two or more bibliographic references imply the same work, we apply a simple clustering method based on matching of reference titles and publication years. 
References to the same work do not contain any term rearrangement and hence can be compared using a string-based similarity metric~\citep{PWH:18}. To compensate for possible typographic differences in reference titles due to the way different authors format references (e.g., with or without quotation marks) and because our parsing method cannot exactly separate titles from extra information provided in the extracted reference text, we rely on a fuzzy matching of reference titles by computing the Levenshtein editing distance~\citep{Levenshtein:65} between them. Two references are clustered together if their Levenshtein ratio is over a given threshold and publication years match. In this case, we create an instance of the class {\texttt{Cluster}} (see Section~\ref{sect:graph}) which integrates similar references.            
Similarly, we can apply a clustering method based on the fuzzy string matching to index labels. However, index terms are much shorter, so clustering only makes sense if the parsing method performs well and the similarity threshold is high. We found clustering useful for indexes that refer to specialized information (e.g., index of ancient author names). In particular, clustering helps to reduce the number of requests to external services when we attempt to link extracted terms with their canonical representation in external databases.  

To enable large-scale automatic bibliographic reference and index term disambiguation, we are interested in services with public APIs (Application Programming Interfaces). API is a protocol that allows a user to query a resource and retrieve data in a machine-readable format. A number of APIs that provide access to collections of published works such as books or scholarly journal articles are available to researchers. Some of them are open to the public, while others are available exclusively to libraries or require subscriptions. Due to the need to disambiguate a large number of extracted references, we are also interested in APIs without rigid constraints on the number of requests. Two global open APIs with unrestricted number of requests that enable search over published works are:
\begin{itemize}
    \item Google Books API, in particular, the {\texttt{Volume}} collection~\cite{googleBooks}, which shares metadata with industry identifiers (DOIs, ISBN) for books or magazines hosted by Google Books. 
    \item Crossref API~\cite{Crossref} which provides metadata with DOIs for 100 million scholarly works.
\end{itemize}
In our pipeline, we disambiguate extracted references by executing HTTP requests to these services that search for works with best matching titles to ours. The responses are given in the form of a set of JSON objects from which we extract the key parameters of the reference, including the title and the publication year. We then check whether (i) the syntactic match  between the returned title and our original title exceeds the similarity threshold and whether (ii) the publication years match. If both conditions hold, we create an instance of the  {\texttt{ExternalPublication}} class (see Section~\ref{sect:graph}) which unambiguously identifies the cited work (via industry identifiers and/or link to the metadata provided by the aforementioned APIs). The experimental evaluation of this process is discussed in Section~\ref{sect:evaluation}. 


Similarly, we disambiguate index terms using two Knowledge Bases (KB):
\begin{itemize}
    \item Wikidata, a free and open KB that acts as central storage for the structured data of projects such as Wikipedia (free encyclopedia), Wikisource (free library), Wiktionary (open dictionary of terms), and others. We rely on its API functionality that enables search for entities using labels and aliases~\cite{wikidata}.
    \item HuCit KB\footnote{\url{https://pypi.org/project/hucitlib}.}~\citep{romanello_pasin_hucitlib} of classical (Greek and Latin) texts, developed with the aim of supporting the automatic extraction of bibliographic references to such texts. As a specialized KB, this resource is particularly useful for the disambiguation of author names and work titles, including their abbreviated versions, that occur in the indexes of ancient names and cited passages.  
\end{itemize}    

We evaluate bibliographic reference and index term disambiguation using the aforementioned methods in Section~\ref{sect:evaluation}.

\section{Data model and Knowledge graph}
\label{sect:graph}

\subsection{Data model}
\label{sect:graph-data}

\begin{figure}
  \centering
  \includegraphics[width=0.95\textwidth]{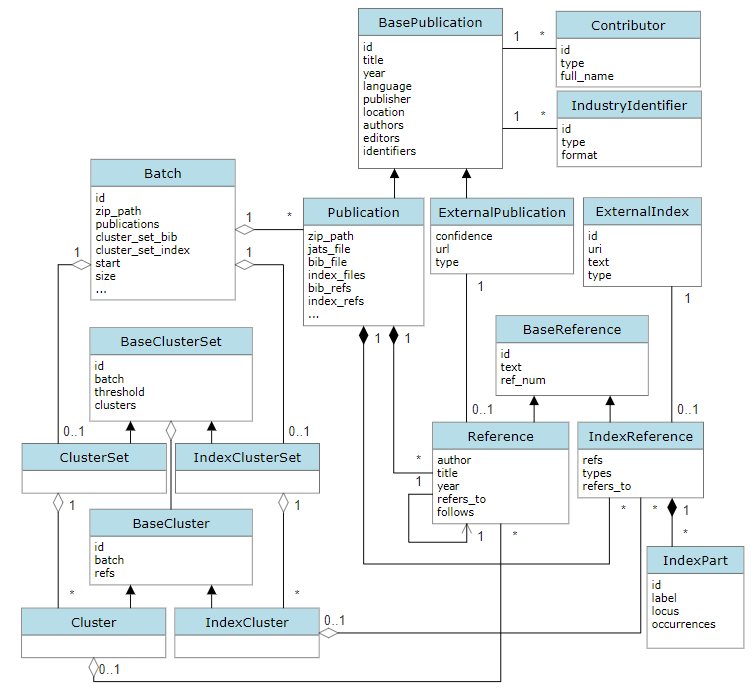}
  \caption{Class diagram of the PDF Parser data model.}
  \label{fig:class-diagram}
\end{figure}

The class diagram in Figure~\ref{fig:class-diagram} shows our model for the representation of extracted data in preparation for the KG construction. As mentioned in Section~\ref{sect:pipeline}, to manage large datasets or publication archives, we split them into batches of a manageable size. All pipeline operations are performed on a single batch, and objects of the {\texttt{Batch}} class are created to represent the extracted data. Each batch records the information about the location of the dataset, its starting point (i.e., list index) and size, as well as a list of publications included to the batch, and, optionally, if the clustering is performed, references to cluster sets that group extracted bibliographic references and index terms.

The publication metadata is accumulated in the {\texttt{BasePublication}} class, which is a common class to define publications from a publisher's archive and external publications added at the bibliographic reference disambiguation step. The base publication object typically would include  title, publication year, language, publisher, authors, editors, and industry identifiers. The object of the derived class {\texttt{Publication}} that represents a publication from the internal dataset additionally records the corresponding location (archive or data folder), relevant back matter file paths, i.e., JATS (BITS) file, PDF bibliography and PDF index files, as well as the lists of extracted bibliographic and index references.    

The latter are represented by instances of the class {\texttt{Reference}} and {\texttt{IndexReference}}, respectively. Both classes extend the abstract class {\texttt{BaseReference}} that keeps the entity text and its order number in the publication's list of references or index terms. The specialization of these classes is in their ability to parse the specific entry text and produce structured representation of either a  bibliographic reference or an index term. The former consists of the author, publication year, and title. It can also keep an optional pointer to a preceding reference to be able to derive author names from it if they are omitted in the reference text. The latter represents an index entry as a set of index parts which includes a label, (optional) locus, and a list of label occurrences (i.e., page numbers). Finally, both types of objects can refer to external resources that disambiguate them, which are modelled by the classes {\texttt{ExternalPublication}} and {\texttt{ExternalIndex}}, respectively. These classes keep links (URIs) to the public resources and the type of the API that supplied them (i.e., Google Books, Crossref, Wikidata, or Hucit). Additionally, to be able to execute meaningful queries over the KG, it is useful to copy some structured data from external objects to our model (e.g., external publication objects provided by Google Books or Crossref APIs allow us to get full contributor names, making it possible, for example, to discover other works of the same authors).         

\subsection{Knowledge graph}
\label{sect:graph-db}

We store the data extracted in the processing pipeline in the form of nodes and relationships using the Neo4j graph database. Node labels in this graph correspond to the classes described above, with the exception of abstract classes and batch objects as they merely serve as helpers in data extraction processes. The relationships originate from the association links between the objects.      

\begin{figure}
    \centering
    \includegraphics[width=1\textwidth]{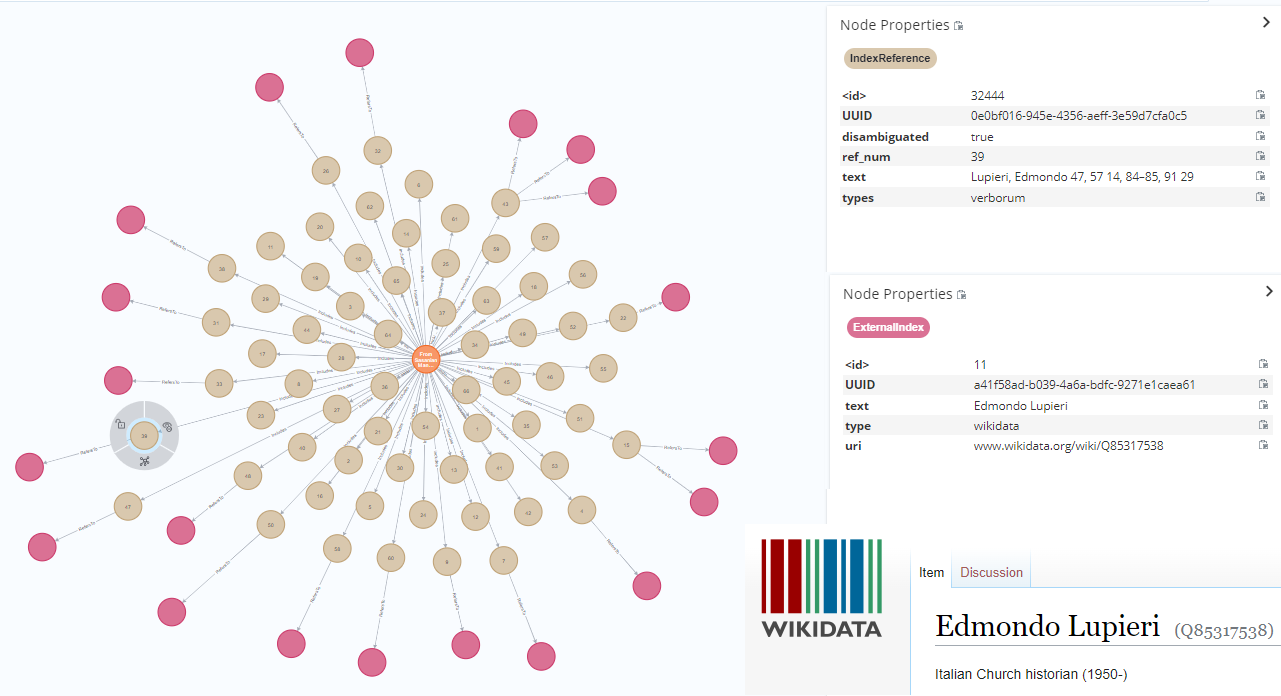}
    \caption{Publication node with related index terms and links.} 
    \label{fig:graph-index}
\end{figure}

Figure~\ref{fig:graph-index} shows a KG fragment that displays a {\texttt{Publication}} node for the book with DOI `B9789004339460' (randomly chosen from the Brill corpus to illustrate our data extraction process), along with a set of relevant index terms represented by the {\texttt{IndexReference}} nodes, with the edges (relations, labelled {\texttt{Includes}}) between them. The general pipeline discussed in Section~\ref{sect:pipeline} discovered the index file in the compressed publication folder, extracted and parsed 66 index entries, out of which 17 were disambiguated via the Wikidata API.  The simple generic index parser we designed failed to parse 25 entries due to the use of the specialized characters from the Brill typeface. The external URLs provided by the Wikidata API spawned the {\texttt{ExternalIndex}} nodes, connected to the corresponding index nodes via the KG edges labelled as {\texttt{RefersTo}}. For example, the selected node corresponds to the entry `Lupieri, Edmondo 9, 80' and resolves to the Wikidata term `Q85317538'.~\footnote{From: \url{https://brill.com/view/book/9789004339460/B9789004339460_018.xml}.}  

\begin{figure}[ht]
    \centering
    \includegraphics[width=1\textwidth]{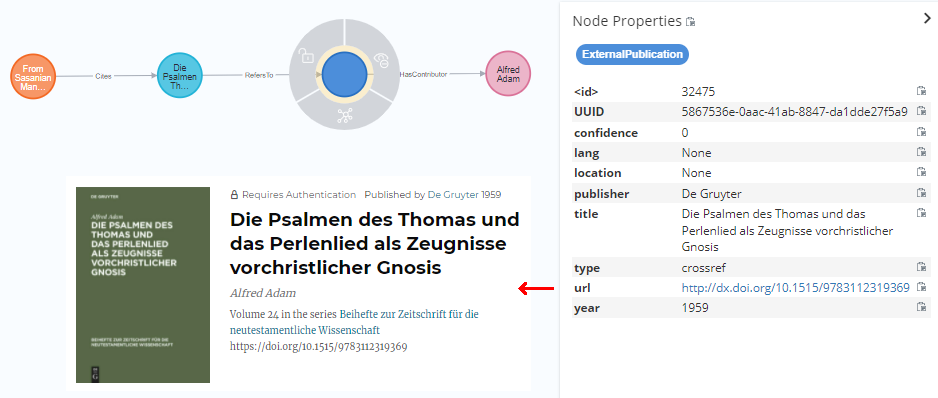}
    \caption{Bibliographic reference node with disambiguation link.} 
    \label{fig:graph-bib}
\end{figure}

Figure~\ref{fig:graph-bib} shows a KG fragment that displays a chain linking the same {\texttt{Publication}} node with a {\texttt{Reference}} node, via the edge labelled {\texttt{Cites}}. The reference node originates from the following text~\footnote{Extracted from the PDF bibliography file at \url{https://brill.com/view/book/9789004339460/B9789004339460_017.xml}.}:
\begin{displayquote}
    \small
    Adam, Alfred. 1959. Die Psalmen des Thomas und das Perlenlied als Zeugnisse vorchristlicher Gnosis, Berlin: Alfred T\"{o}pelmann.
\end{displayquote}
The next node in the line is the {\texttt{ExternalPublication}} node that disambiguates this reference via the Crossref API. Finally, the last shown node is the {\texttt{Contributor}} node that provides the name and surname of the cited work's author, extracted from the Crossref's record.


\subsection{Implementation and dissemination}
\label{sect:graph-code}

The source code for our information extraction and KG construction software is openly available~\footnote{\url{ https://github.com/kiem-group/pdfParser}.}. The Python package provides a proof-of-concept implementation of the described pipeline (see Section~\ref{sect:pipeline}) that expects as input a path to a folder with compressed publication archive (see Section~\ref{sect:content-brill}) and a URI to the Neo4j database instance to store the KG. Besides the implementation of the data model (see Section~\ref{sect:graph-data}), the toolset includes a {\texttt{DBConnector}} class which provides CRUD (Create, Read, Update, Delete) operations on the KG and useful methods for data refinement such as cluster or external link node merging. These operations can be exposed via a REST API to other systems to retrieve or manipulate the acquired data, in particular, enable  useful queries for data analysis (which, however, is out of the scope of the current project). We also illustrate, via a series of tests, how the pipeline can be customized with different parsers and disambiguation services, and experiment with alternative serialization methods such as translation of our  data model (see Section~\ref{sect:graph-data}) into formats defined by SPAR (Semantic Publishing and Referencing) ontologies. 
The SPAR ontologies form a suite of modules for the creation of comprehensive machine-readable RDF (Resource Description Framework) metadata for every aspect of semantic publishing and referencing such as document description, bibliographic resource identifiers, types of citations and related contexts, bibliographic references, document parts, agents' roles and contributions, and more~\citep{vrandecic_spar_2018}.
We enable partial mapping of our data model into SPAR RDF. Listing~\ref{lst:spar-map} illustrates how a publication with corresponding automatically extracted bibliographic references can be represented in the OpenCitations Data Model (OCDM)~\citep{pan_opencitations_2020}. 

\begin{lstlisting}[caption={SPAR bibliographic references}, label={lst:spar-map}]
    @prefix : <http://brill.com/kiem/> .
    @prefix biro: <http://purl.org/spar/biro> .
    @prefix co: <http://purl.org/co/> .
    @prefix frbr: <http://purl.org/vocab/frbr/core#> .
    @prefix c4o: <http://purl.org/spar/c4o> .
    <http://dx.doi.org/10.1163/9789004261648> frbr:part :reference-list .
    :ref-001 a biro:BibliographicReference ;
        c4o:hasContent "Augustine Confessiones. Edited by Martin Skutella. 
        Leipzig: Teubner, 1934. " .
    :ref-002 a biro:BibliographicReference ;
        c4o:hasContent "Chaldaean Oracles Confessions. Translated by Henry 
        Chadwick. Oxford: Oxford University " .
    ...       
\end{lstlisting}

The OCDM is the most notable realization of the SPAR general guidelines used for storing data in OpenCitations datasets~\citep{peroni_opencitations_2020}. One of such datasets, COCI~\citep{coci}, the OpenCitations Index of Crossref open DOI-to-DOI citations, is an RDF dataset containing details of all the citations that are specified by the open references to DOI-identified works present in Crossref. 
Table~\ref{tab:ref-num} compares the number of extracted and disambiguated references for five randomly chosen publications from the Brill catalogue with the number of DOI-to-DOI citations provided by the COCI public API~\footnote{\url{http://opencitations.net/index/coci/api/v1}.}. The significant gap between the number of actually cited works in a book, or even between the number of works we were able to disambiguate (associate with their unique identifiers, DOI and/or ISBN, via Crossref and Google Books APIs), suggests that our tool could be employed by other agents, publishers or authors, for augmenting existing datasets with new relationships.

\begin{table}[ht]
    \centering
    \footnotesize
    \begin{tabular}{r|r|r|r|r|r}
        \# refs / DOI    & 9789004180994 & 9789004189829 & 9789004231283 & 9789004257788 & 9789004277151 \\ 
        \hline
        Extracted     & 335           & 325           & 365           & 366           & 220           \\
        Disambiguated & 145           & 96            & 219           & 138           & 94            \\
        COCI          & 9             & 40            & 73            & 6             & 7             \\
        \hline
    \end{tabular}
    \caption{Number of extracted references vs number of COCI DOI-to-DOI citations.}
    \label{tab:ref-num}
\end{table}


\section{Evaluation}
\label{sect:evaluation}

In this section, we reflect on the quality of extracted data by providing some experimental evaluation of the involved pipeline steps.   

\subsection{Quality of PDF parsing}
The first critical step in data extraction process is splitting the back matter file into individual bibliographic and index references, i.e., a heuristic method relying on the alignment of reference text in a PDF file (see Section~\ref{sect:pipeline-pdf}). One way to evaluate the quality of this method would be to compare the extracted reference text with ``ground truth'' reference text in a labelled dataset. We employed this method in unit tests on selected references, but, unfortunately, we did not have access to a sufficiently large dataset of this kind. Nevertheless, the evaluation of this step is subsumed to the evaluation of the disambiguation steps as the human curator evaluating the correctness of external links could also judge the quality of the randomly selected references. Out of a sample of 500 text fragments assumed to be bibliographic references, 495 were valid and complete references, and
\begin{itemize}
    \item 2 entries contained other type of information; they were dismissed by the bibliographic reference parser as no publication year was found in these fragments;
    \item 3 entries contained incomplete reference text: 2 were lacking the tail part yet contained enough information to identify the referred works, and 1 lacked the head part containing authors and publication year.           
\end{itemize}
The aforementioned observations make us confident that our layout-based reference extraction method works very well given a PDF file with list of references, demonstrating almost 99\% accuracy on randomly selected subsets from Brill's corpus. 

The risk of KG contamination still cannot be excluded if the same approach was to be applied to a PDF containing other type of data (e.g., book body chapter) or unusually structured bibliography file that alternates references with paragraphs of other content. Hence, we parsed only back matter files which we were sure contain the right information based on the presence of keywords ``bibliography'' and ``index'' in their titles. Given this constraint, in the Brill's corpus of 1899 publications, the pipeline processed 650 bibliography and 1804 index files.
 
\subsection{Quality of clustering}
For the evaluation of the clustering method we used a ground truth dataset ~\citep{matteo_romanello_2022_7078762}, consisting of 3579 pairs of bibliographic references.
For each labelled pair, a manually assigned score indicates 
whether the two references are referring to the same bibliographic entity or not (1.0 = true, 0.0 = false, 0.5 = partly), e.g.:
\begin{displayquote}
    \small
    C. Lane, Venise, une R\'epublique maritime, Paris, 1988, p. 344; \\
    Lane, F.: Venise, une r\'epublique maritime. Paris 1985. p. 69. \\
    Score: 0.0 \\
    ...\\
    C. Lane, Venise, une R\'epublique maritime, Paris, 1988, p. 344; \\
    Lane, Fr\'ed\'eric Chapin. Venise: une r\'epublique maritime, pr\'eface de Fernand Braudel; trail, de l'am\'ericain par Yannick Bourdoiseau et Marie Ymonel. Paris, Flammarion. 1988. \\
    Score: 1.0 
\end{displayquote}

\begin{figure}
    \centering
    \begin{tikzpicture}
        \begin{axis}[
            title={Accuracy vs similarity threshold},
            xlabel={Threshold},
            ylabel={Accuracy},
            scale only axis,
            width=6cm,
            height=3cm,
            xmin=0.45, xmax=1.05,
            ymin=0.55, ymax=0.95,
            xtick={0.5,0.6,0.7,0.8,0.9,1.0},
            ytick={0.6, 0.7,0.8,0.9,0.9},
            ymajorgrids=true,
            grid style=dashed,
            typeset ticklabels with strut,
            legend pos=outer north east,
            legend cell align=left,
            smooth
        ]
        \addplot[color=blue, mark=*, line width = 1.2pt,]
            coordinates {(0.5,0.8868)(0.55,0.8982)(0.6,0.8982)(0.65,0.9025)(0.7,0.9083)(0.75,0.9075)(0.8,0.905)(0.85,0.9022)(0.9,0.9016)(0.95,0.9033)(1,0.8748)};
       \addplot[color=red, mark=square, line width = 1.2pt,]
            coordinates {(0.5,0.6365)(0.55,0.6859)(0.6,0.7019)(0.65,0.7270)(0.7,0.7541)(0.75,0.7784)(0.8,0.7781)(0.85,0.7773)(0.9,0.7790)(0.95,0.7887)(1,0.8041)};    
        \legend {
            Publication year + title,
            Title
        }    
        \end{axis}
    \end{tikzpicture}
    \caption{Evaluation of the bibliographic reference clustering.}
    \label{fig:eval-lev}
\end{figure}
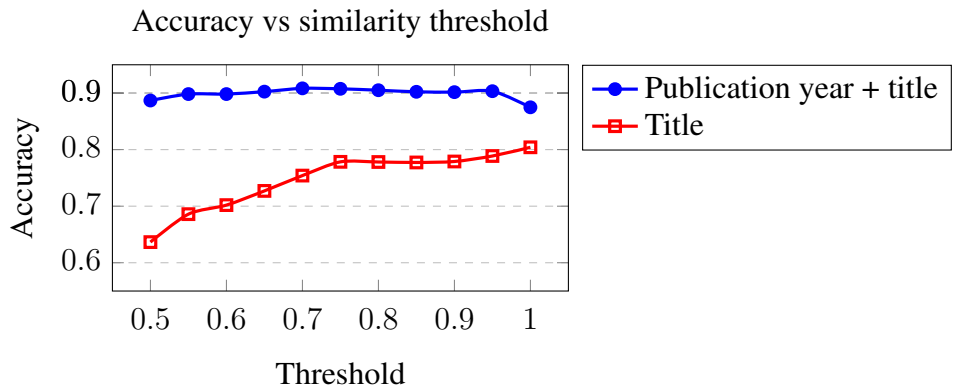

We parsed the given pairs of text references and computed the Levenshtein similarity ratio between their titles (more specifically, the shortest title and the part of the longer title of the same length). Two references are clustered together if their publication years match and the similarity score exceeds a given threshold. The first plot in Figure~\ref{fig:eval-lev} shows the accuracy of the similarity-based clustering for various thresholds, reaching 90.8\% accuracy for the value of $0.7$. 

The close accuracy values for various thresholds indicate that the decision in this artificial dataset significantly depends on the matching of the publication years. Indeed, by matching publication years alone, we yield the $82\%$ accuracy. Obviously, in large publication archives, title matching will have the primary role in clustering references to the same work. To choose the best similarity threshold, we therefore repeated this experiment matching publication titles alone. The second plot in Figure~\ref{fig:eval-lev} shows that, for the threshold values above $0.75$, the accuracy is close to $80\%.$   

Based on this evaluation, we chose the default threshold of $0.75$ for bibliographic reference title matching in disambiguation methods of our pipeline.   

\subsection{Quality of disambiguation}

To evaluate the quality of disambiguation step, we randomly selected 500 extracted references from Brill's corpus, and generated a spreadsheet that associates the reference text with URLs supplied by the Google Books and Crossref APIs in response to the disambiguation requests in the pipeline. We then asked a human expert, a Brill's employee, to evaluate whether the URLs refer to the correct work, with score being 0 for a wrong or absent link, 1 for the correct link, and 0.5 for a partially correct link, i.e., a URL of the book collection containing the cited work or a different edition of the cited work. Moreover, for entries with incorrect or missing links, the expert was asked to manually locate publicly available link matching the work in the reference. Similarly, a spreadsheet containing 500 randomly selected index references was generated together with Wikidata responses disambiguating labels from these entries. In our early experiments, we observed that the Wikidata API~\footnote{\url{https://www.wikidata.org/w/api.php?action=wbsearchentities&search}.} often does not produce results for queries containing author surname followed by the comma-separated name(s) as opposed to the name(s) followed by the surname (e.g., ``Gournay, Marie le Jars de'' vs ``Marie le Jars de Gournay''). Adding an extra query with transformed search label for indexes with a comma after the first term helped us to compensate for this effect and acquire identifiers for more entries. 

Our disambiguation method resolved 209 entries for 500 bibliographic references, and the human expert assigned the total score of 169 points to their correctness; this allows us to estimate its precision as $169/209$, or $80.8\%.$ At the same time, the human expert was able to find 149 extra links in addition to the 209 we filled in automatically, increasing the number of disambiguated references to 358. On this ground, we estimate the recall of our method as $209/358$ or $58.4\%.$ 
The analogous approach for the index disambiguation yielded the score of 259 for the total of 274 automatically discovered  Wikidata identifiers, setting precision to $94.5\%.$ With the overall total of 414 links, automatically and manually located, the recall is estimated to be at $66\%.$~\footnote{The complete evaluation tables are available at \url{https://github.com/kiem-group/pdfParser/tree/main/data_test}.}

We did not evaluate Hucit-based disambiguation in the same style since it is a highly specialized database that searches for exact terms and, considering that we do not retain context information, any response, if present, is likely correct. The coverage hugely depends on the index content and hence is not representative either. Moreover, its API generates much slower responses, so this disambiguation service is recommended for targeted disambiguation, when a human supervisor considers its use more promising than the generic Wikidata search.

\section{Related Work}
\label{sect:related-work}

Our work relates to three distinct areas of research: a) the construction of knowledge graphs, b) the extraction of bibliographic information from AHSS publications and c) the processing of indexes.

\subsection{Knowledge Graphs}

The development of KGs in the publishing sector is part of the ongoing transformation of publications from document-centred objects to structured, interlinked knowledge representations. KGs such as Microsoft Academic Knowledge Graph \citep{ghidini_microsoft_2019}, SciGraph \citep{yaman_interlinking_2019}, the Literature Graph \citep{ammar_construction_2018} and the Semantic Scholar Open Research Corpus \citep{lo_s2orc_2020} underpin the search functionalities of academic search engines like Google Scholar, Microsoft Academic and Semantic Scholar. AHSS disciplines, however, tend to be underrepresented and poorly covered in such KGs which are developed with a focus on STM disciplines. Within this landscape, the open KG of Wikidata represents quite an exception as several resources in the Humanities (e.g., gazetteers, catalogues) have been linked to entries in Wikidata, thus making it a useful resource when extracting and linking structured knowledge from publications in this area \citep{haslhofer_knowledge_2018}.  
The High Integration of Research Monographs in the European Open Science (HIRMEOS) project~\cite{HIRMEOS-2017} is to address the particularities of academic monographs in the AHSS domain and foster integration  of monographs into the European Open Science Cloud. 

\subsection{Citation Mining}

In this paper, we approached the extraction of bibliographic references by parsing the bibliography section of each book, after having it located by applying some heuristics on the available book-level metadata. This approach, however, is not viable when structural metadata about the publication are lacking, or when references are not grouped in a dedicated section (such is the case with journal articles). An alternative to the parsing of bibliography sections are end-to-end approaches which locate references within the full-text of publications, and successively segment them into their components. Existing services that implement such end-to-end extraction of references are GROBID \citep{lopez_grobid_2009}, CERMINE \citep{tkaczyk_cermine_2015}, BILBO \citep{kim_automatic_2011} and EXCITE \citep{hosseini_excite_2019-1}, with the last two having a specific focus on AHSS publications. 
While these services are able to recognize references in mixed documents, their evaluation on our sample PDF documents containing exclusively bibliographic references revealed that the basic heuristic-based reference splitting procedure we proposed performs significantly better.     

While the large majority of work in this area dealt with references to secondary literature, the extraction of references to primary sources was investigated, in particular with respect to archival references \citep{rodrigues_alves_deep_2018} and canonical references to Greek and Latin literary works \citep{romanello_index_2015}. 


\subsection{Index Processing}

When it comes to the processing of indexes, we should distinguish between a) processing aimed at the semi-automatic creation of indexes and b) processing existing indexes (i.e., parsing) with the aim of exploiting the structured information they contain.

With respect to the creation of indexes, the landscape is dominated by commercial indexing software (e.g., CINDEX, Macrex, Sky Index) that automate certain indexing-related tasks -- such as sorting, formatting, cross-referencing -- while the human in the loop is still responsible for the main intellectual work entailed by creation of an index. \cite{reitz_experiments_2019}, in collaboration with the publisher De Gruyter, has recently showed how the creation of an index locorum containing several thousands of entries can be largely automated by using open citation mining software.

Researchers, however, have also recognized the value of existing indexes as sources of structured, human-curated information. Recent work by \cite{blidstein_towards_2022} provides a fitting example of how indexes, and \textit{indices locorum} in particular -- can provide extremely valuable data for citation analysis.
By parsing such indexes, they distilled information about cited primary sources within a corpus of scholarly books on ancient Mediterranean religion and culture. 
They were able to explore how various network types capture distinct aspects of published scholarship thanks to this corpus, which also allowed for systematic analysis and comparison of various citation network types.
Furthermore, parsing of digitized indexes was employed for the construction of controlled vocabularies \citep{piotrowski_harvesting_2012}, gazetteers of historical places \citep{piotrowski_leveraging_2010} and literary authors and works \citep{romanello_when_2009-1}, for improving access to historical archives \citep{colavizza_index-driven_2019}, for building character networks from literary texts \citep{rochat_character_2014}, and for reconstructing the ontological relations implied by hierarchical relations between index terms \citep{li_extracting_2019}. 


\section{Conclusion}

In this work, we proposed an open and collaborative approach to facilitating the indexation of AHSS literature in modern scientific search engines. We openly released a software implementing a pipeline able to extract structured information from the back matter of books, specifically references and indexes. The pipeline further implements disambiguation and normalization routines on the extracted information, and constructs a knowledge graph using the SPAR and OpenCitations data models. Our work is aimed at supporting small and medium enterprise (SME) publishers in the arts, humanities and social sciences (AHSS). We seek to provide them with a means of contributing their own publication knowledge graphs in the open domain. To exemplify and evaluate the proposed pipeline, we applied it on Brill's Classics corpus, with promising results.

We identify several directions for future work. Firstly, our proposed pipeline implementation can be improved in a variety of ways, for example by expanding the coverage of data formats and contents it can handle, and the methods used for PDF parsing, clustering or entity disambiguation. We believe this will happen only if the pipeline is adopted by a community of users, primarily AHSS SME publishers. This is why another important direction for future work includes the dissemination of our project results and the maintenance of the software we released. Lastly, more effort is needed to fully integrate the proposed pipeline within the existing and growing open science ecosystem. Making sure that the extracted information is linked to mainstream open authority records, as well as ingested in OpenCitations is of critical importance to the success of the proposed approach. Our contribution here made the first, but not yet all the steps in this direction.

Ideally, in the future we hope to see a wide adoption of the approach we propose by AHSS publishers, so that the software we released will become a live codebase with a diverse set of users and contributors. We believe that, with an open source, collaborative approach, we can significantly improve the indexing and discoverability of AHSS literature.

\section*{Code and data availability}

The source code of the proposed information extraction and knowledge graph construction software is openly available: \url{ https://github.com/kiem-group/pdfParser}.

\section*{Acknowledgements}

The authors would like to thank Anna van den Bosch for her  help with the extracted data quality evaluation, and Silvio Peroni for his insights on the use of SPAR ontologies and OpenCitations data model.  

GC and NK acknowledge the support of the Dutch Research Council (NWO), grant number KIEM.K20.01.137. 

\bibliographystyle{plainnat}
\bibliography{main}

\appendix\footnotesize


\end{document}